\pdfoutput=1 
\documentclass[orivec]{llncs}

\usepackage{epsfig}
\usepackage{xspace}
\usepackage{wrapfig}
\usepackage{listings}
\usepackage{latexsym,amssymb}
\usepackage{amsmath}
\usepackage{alltt}
\usepackage[all]{xy}
\usepackage{algorithm2e}
\usepackage{algorithmicx}
\usepackage{multirow}
\usepackage{color} 
\usepackage{graphics}
\usepackage{graphicx}
\usepackage[draft]{fixme}
\usepackage{booktabs}
\usepackage{bold-extra}
\usepackage{hyperref}
\usepackage{listings}
\lstdefinestyle{PrologStyle} {
  language=Prolog,
  framerule=0pt,
  basicstyle=\footnotesize\ttfamily,
  commentstyle=\color{black},
  keywordstyle=\footnotesize\ttfamily\bfseries,
  stringstyle=\footnotesize\ttfamily,
  showstringspaces = false,
  escapechar=@,
  mathescape=true,
  extendedchars=true,%
  lineskip=0ex,%
  framerule=0pt,
  numbersep=2pt,
  numberstyle=\tiny,
  deletekeywords={length,ef}, 
  morekeywords={set_field,get_field,return,async,get,new_object},
}

\include{costa-defs}

\newcommand{\maxcard}{C^{max}}
\newcommand{\mincard}{C^{min}}
\newcommand{\initcontexts}[1]{\mathcal{I}^{#1}}
\newcommand{\inittasks}{\mathcal{T}_{ini}}

\newcommand{\buffer}{\mathit{loc}}
\newcommand{\queue}[0]{\C{Q}}

\newcommand{\macrostep}[1]{\stackrel{#1}{\longrightarrow}}

\newcommand{\C}[1]{{\cal #1}}

\newcommand{\namemethod}{m}
\newcommand{\mapping}{l}
\newcommand{\heap}{h}

\newcommand{\objid}[0]{\ensuremath{\mathit{o}}\xspace}

\newcommand{\tkid}[0]{\ensuremath{\mathit{tk}}\xspace}

\newcommand{\lst}[1]{\lstinline!#1!}

\newcommand{\Await}[0]{\ensuremath{\mbox{\bf \lstinline!await!}}\xspace}
\newcommand{\get}[1]{\ensuremath{#1.\mbox{\bf \lstinline!get!}}\xspace}
\newcommand{\Get}[1]{\ensuremath{\mbox{\bf \lstinline!get!}}\xspace}
\newcommand{\newb}{\ensuremath{\mbox{\lstinline!new!}}\xspace}
\newcommand{\ifte}{\mbox{\lstinline!if!}\xspace}
\newcommand{\iftethen}{{\ensuremath{\mbox{\lstinline!then!}}\xspace}}
\newcommand{\ifteelse}{{\ensuremath{\mbox{\lstinline!else!}}\xspace}}
\newcommand{\while}{\mbox{\lstinline!while!}\xspace}
\newcommand{\whilebody}{\mbox{\lstinline!do!}\xspace}
\newcommand{\return}{\mbox{\lstinline!return!}\xspace}
\newcommand{\await}[1]{\mbox{\bf \lstinline!await!}~#1?\xspace}
\newcommand{\returnval}[1]{\mbox{\lstinline!return #1!}\xspace}
\newcommand{\body}[1]{\ensuremath{\mathit{body}(#1)}}

\newcommand{\extend}[1]{S}
\newcommand{\exec}{{\it exec}}

\newcommand{\anndep}[1]{\ensuremath{\xrightarrow{#1}}}

\newcommand{\abstkid}[0]{\ensuremath{\mathit{tk}}\xspace}

\newcommand{\taskpp}[2]{\ensuremath{{#1}{:}{#2}}}

\begin{document}

\title{On the Generation of Initial Contexts for Effective Deadlock
  Detection\\ (Extended Abstract)\vspace{-.3cm}\thanks{This work was
    funded partially by the Spanish MINECO project
    TIN2015-69175-C4-2-R, and by the CM project S2013/ICE-3006.}}
\author{\vspace{-.0cm}Elvira Albert \and Miguel G\'omez-Zamalloa \and Miguel Isabel}

\institute{
Complutense University of Madrid (UCM),  Spain
\vspace{-.1cm}}

\setcounter{page}{1}

\maketitle

\label{firstpage}

\begin{abstract}
  It has been recently proposed that testing based on symbolic
  execution can be used in conjunction with static deadlock analysis
  to define a deadlock detection framework that: (i) can show deadlock
  presence, in that case a concrete test-case and trace are obtained,
  and (ii) can also prove deadlock freedom.
 Such symbolic execution starts from an 
  \emph{initial distributed context}, i.e., a set of locations and their
  initial tasks.
%
  Considering all possibilities results in a combinatorial explosion
  on the different distributed contexts that must be considered. This
  paper proposes a technique to effectively generate initial contexts
  that can lead to deadlock, using the possible conflicting task
  interactions identified by static analysis, discarding other
  distributed contexts that cannot lead to deadlock.
  The proposed technique has been integrated in the above-mentioned
  deadlock detection framework hence enabling it  to analyze systems
  without the need of any user supplied initial context.
%
\end{abstract}%


\section{Motivation}\label{sec:intro}

\begin{figure}[t]
\begin{center}
\begin{minipage}{8.5cm}
\end{minipage}
\begin{tabular}{ll}

\begin{lstlisting}[name=smallexamples]
main(){$\label{pp:main}$
    DB db = new DB();$\label{pp:newdb}$
    $\label{pp:neww}$Worker w = new Worker();
    $\label{pp:asyncr}$db!register(w);
    $\label{pp:asyncw}$w!work(db);}

class Worker{
  Data data;
  int work(DB db){$\label{pp:workinit}$
    Fut$\langle$Data$\rangle$ f = db!getData(this); $\label{pp:fgetdata}$
    $\label{pp:work}$data = f.get;
    return 0;
  }
  int ping(int n){return n;}$\label{pp:ping}$ 
}// end of class Worker

class DB{
  Data data = ...;
  Worker client = null;
  int connected = 1;
\end{lstlisting}
~&~
\begin{lstlisting}[name=smallexamples]
  int makesConnection(){$\label{pp:makesinit}$
      connected = 3;$\label{pp:makestrue}$ 
      return connected;
  } 
  int register(Worker w){$\label{pp:reginit}$
      connected = 5; $\label{pp:connected1}$
      Fut$\langle$int$\rangle$ g = this!getData();
      $\label{pp:await} $await g?;
      if (connected > 0){ 
         connected = connected - 1; $\label{pp:connected}$
         Fut$\langle$int$\rangle$ f = w!ping(5);$\label{pp:fping}$
         if (f.get == 5) client = w;$\label{pp:register}$
      } 
      return 0;
  } 
  Data getData(Worker w){$\label{pp:getD}$
    if (client == w) return data;
    else return null;
  }
}// end of class DB
\end{lstlisting}
\end{tabular}\vspace{.1cm}

\end{center}
\caption{Working example. Communication protocol between a DB and a worker}
\label{fig:examples}
\end{figure}

Deadlocks are one of the most common programming errors and they are
therefore one of the main targets of verification and testing tools.
We consider a distributed programming model with explicit
\emph{locations} (or distributed nodes) and \emph{asynchronous} tasks
that may be spawned and awaited among locations. Each location
represents a processor with a procedure stack and an unordered queue
of pending tasks. Initially all processors are idle. When an idle
processor's task queue is non-empty, some task is selected for
execution, this selection is non-deterministic.
Let us see now our motivating example in Figure~\ref{fig:examples}
which simulates a simple communication protocol between a database and
a worker.  Our implementation has the \lst{main} method, and two
classes \lst{Worker} and \lst{DB} implementing the worker and the
database, respectively.  The \lst{main} method creates two distributed
locations: the database and the worker, and (asynchronously) invokes
methods \lst{register} and \lst{work} on each of them,
respectively. The \lst{work} method of a worker simply accesses the
database (invoking asynchronously method \lst{getData}) and then
\emph{blocks} until it gets the result, which is assigned to its
\lst{data} field. The instruction \lst{get} blocks the execution in
the current location until the awaited task has terminated. We use
future variables \cite{deboer07esop-short,flanagan95:futures} to detect the
termination of asynchronous tasks.
The \lst{register} method of the database makes a call to
\lst{getData} and waits for its execution.  Once it has finished, it
checks if the number of possible connections is bigger than 0. In that
case \lst{connected} is decreased by one, and the database makes sure
that the worker is online. This is done by invoking asynchronously
method \lst{ping} with a concrete value and blocking until it gets the
result with the same value. Then, the database registers the provided
worker reference storing it in its \lst{client} field.
Method \lst{getData} of the database returns its \lst{data} field if
the caller worker is registered, otherwise it returns
\lst{null}. Finally, method \lst{makesConnection} sets the field
\lst{connected} to $3$.
Depending on the sequence of interleavings, the
execution of this program can finish: (1) as one would expect, i.e.,
with \lst{worker.data = db.data}, (2) with \lst{w.data = null} if
\lst{getData} is executed before the assignment at line
\ref{pp:register}, or, (3) in a deadlock.


We have recently proposed a deadlock detection
framework~\cite{AlbertGI16b-short,AlbertGI17_techrep} that combines static
analysis and symbolic execution based
testing~\cite{DBLP:conf/hvc/AgarwalWS05,AlbertGI16b-short,DBLP:conf/icst/ChristakisGS13,DBLP:conf/sigsoft/JoshiNSG10}. The
deadlock analysis (for instance, \cite{FloresAG13-short}) is first used to obtain
descriptions of potential deadlock cycles which are then used to guide
the testing process. 
The resulting deadlock detection framework hence can: (i) show
deadlock presence, in which case a concrete test-case and trace are
obtained, and (ii) prove deadlock freedom (up to the symbolic
execution exploration limit).
%
%
However, the symbolic execution phase needs to start from a concrete
initial distributed context, i.e., a set of locations and their
initial tasks.
%
In our example, such an initial context is provided by the \lst{main}
method, which creates a \lst{Database} and a \lst{Worker} location,
and schedules a \lst{work} task on the worker with the database as
parameter, and, a \lst{register} task on the database with the worker
as parameter. This is however only one out of the possible contexts,
and, of course, it could be the case that it does not expose an error
that occurs in other contexts (for instance, it does not manifest any
deadlock). This clearly limits the framework potential.


A fundamental challenge for a symbolic execution framework of
distributed programs is to automatically and systematically generate
\emph{relevant} distributed contexts for the type of error that it
aims at detecting. This would allow for instance applying symbolic
execution for system and integration testing.
The generation of relevant contexts involves two challenging aspects:
(1) A first challenge is related to the elimination of redundant
(useless) contexts.  Observe that there is a combinatorial explosion
on the different possible distributed contexts that can be generated
when one considers all possible types and number of distributed
locations and tasks within them.  Therefore, it is crucial to provide
the \emph{minimal} set of initial contexts that contains only one
representative of  equivalent contexts. 
%
(2) For the particular type of error that one aims at detecting, an
additional challenge is to be able to only generate initial contexts
in which the error can occur.  
%
In the case of generating initial contexts for deadlock detection in
our working example, this would mean generating for instance, a context
with a database location and some worker location with a scheduled
\lst{work} task and a \lst{register} task on the database for it,
i.e., the context created by the \lst{main} method. For instance, contexts
that do not include both tasks would be useless for deadlock
detection. Let us observe that if the assignment at
Line~\ref{pp:makestrue} is changed to assign $0$, then the initial
contexts must also include a \lst{makesConnection} task, otherwise no
deadlock will be produced.
Interestingly, deadlock analyses provide
\cite{FloresAG13-short,DBLP:conf/concur/GiachinoKL14,DBLP:journals/sosym/GiachinoLL16}
potential \emph{deadlock cycles} which contain the possibly
conflicting task interactions that can lead to deadlock. This
information will be used to help our framework anticipate this
information and discard initial distributed contexts that cannot lead
to deadlock from the beginning.  Briefly, the main contributions of
this work are twofold:
\begin{itemize}
\item We introduce the concept of \emph{minimal} set of initial
  contexts and extend a static testing framework to automatically and
  systematically generate them.
\item We present a deadlock-guided approach to effectively generate
  initial contexts for deadlock detection.
\end{itemize}
In an extended version of this work, we will validate experimentally
our proposal and prove its soundness formally.

\section{Asynchronous Programs}\label{sec:lang}
A program consists of a
set of classes that define the types of locations, each of them
defines a set of fields and methods of the form $M{:}{:}{=} T~
m(\bar{T}~ \bar{x}) \{s\}$, where statements $s$ take the form $\sf s
{:}{:}{=} s; s \mid x{=}e \mid
\ifte~\mathit{e}~\iftethen~s~\ifteelse~s \mid
\while~\mathit{e}~\whilebody~\mathit{s} \mid \return~x; \mid
b{=}\newb~T({\bar{z}})\mid f=x~{\bold !}~m({\bar{z}}) \mid \await{f} \mid
x=\get{f}$.
%
%
 Syntactically, a location will
 therefore be similar to a \emph{concurrent object} that can be
 dynamically created using the instruction \newb $\sf T(\bar{z})$. 
The
declaration of a future variable is as follows $\sf Fut \langle T \rangle~f$, where
$\sf T$ is the type of the result $\sf r$, it adds a new future
variable to the state.  
Instruction $\sf f=x~{\bold !}~m({\bar{z}})$ spawns a new task (instance of
method {\sf m}) 
and it is set to the future {\sf f} in the state. 
Instruction \await{f} allows non-blocking
synchronization.  If the future variable {\sf f} we are awaiting for
points to a finished task, then the \Await{} can be
completed. Otherwise the task yields the lock so that any other task
of the same location can take it.
On the other hand, instruction $\get{f}$ allows blocking
synchronization. It waits for the future variable without yielding the
lock, i.e., it blocks the execution of the location until the task
that is awaiting is finished. Then, when the future is ready, it
retrieves the result and allows continuing the execution.  This
instruction introduces possible deadlocks in the program, as two tasks
can be awaiting for termination of tasks on each other's locations.
Finally, instruction $\return~x;$ releases the lock that will never be
taken again by that task. Consequently, that task is \emph{finished}
and removed from the task queue.
All statements of a task takes place serially (without interleaving
with any other task) until it gets to a \return or \await{f}
instruction. Then, the processor becomes idle again, chooses
non-deterministically the next pending task, and so on.

A \emph{program state} or \emph{configuration} is a set of locations
$\{\mathit{loc}_0,...,\mathit{loc}_n\}$.
 A \emph{location} is a term $\buffer(\objid,\tkid,\heap,\queue)$
 where $\objid$ is the location identifier,
$\tkid$ is the identifier of the \emph{active task} that holds the
location's lock or $\bot$ if the location's lock is free, $h$ is its
local heap, and $\queue$ is the set of tasks in the location.
A \emph{task} is a term $tsk(\tkid,\namemethod,\mapping,s)$ where
$\tkid$ is a unique task identifier, $\namemethod$ is the method name
executing in the task, $\mapping$ is a mapping from local variables to
their values, and $s$ is the sequence of instructions to be executed. 
We assume that the execution starts from a $main$ method without
parameters. The initial state is
$\extend{\initstate}{=}\{\buffer(0,0,\bot,\{tsk(0,main,\mapping,\body{main})\}$
with an initial location with identifier $0$ executing task $0$,
%
maps local variables to their initial values, and $\body{m}$ is the
sequence of instructions in method $m$ and $ini(main)$ is the initial
program point in method $m$. From now on, we represent the state as a
Prolog list, and we write $[x \mapsto v]$ to denote $h(x)=v$
(resp. $l(x)=v$), that is, field $x$ in the heap $h$ (resp. local
variable $x$ in the mapping $l$) takes the value $v$.

 In what follows, a \emph{derivation} or \emph{execution}
\cite{DBLP:conf/fase/SenA06} 
is a sequence of states 
 $S_0 \macrostep{o_1.t_1} ...
 \macrostep{o_{n}.t_{n}} S_n $, where $S_i
\macrostep{o_i.t_i} S_{i+1}$ denotes the execution of task $t_i$ in
location $o_i \in S_i$. The derivation is
\emph{complete} if $S_0$ is the initial state and $\nexists~loc(\objid,\_,\_,\{\tkid\} \cup\queue) \in
S_n$ such that
$S_{n} \macrostep{\objid.\tkid} S_{n+1}$ and $S_n \neq S_{n+1}$. 
Given a state $S$, $\exec(S)$ denotes the set
of all possible complete executions starting at $S$. 

\section{Specifying and Generating Initial Contexts}
\label{sec:initialcontexts}

In our asynchronous programs, the most \emph{general} initial contexts
consist of sets of locations with \emph{free} variables in their fields, and
initial tasks in each location queue with \emph{free} variables as
parameters, i.e., neither the fields nor the parameters have concrete
values.  A first approach to systematically generate initial contexts
could consist in generating, on backtracking, all possible multisets
of initial tasks (method names), and for each one, generate all
aliasing combinations with the locations of the tasks belonging to the
same type of location. They are multisets because there can be
multiple occurrences of the same task.  To guarantee termination of
this process we need to impose some limit in the generation of the
multisets. For this, we could simply set a limit on the multiset
global size. However it would be more reasonable and useful to set a
limit on the maximum cardinality of each element in the multiset. To
allow further flexibility, let us also set a limit on the minimum
cardinality of each element. For instance, if we have a program with just one
location type $A$ with just one method $m$, and we set $1$ and $2$ as
the minimum and maximum cardinalities respectively, then there are two
possible multisets, namely, $\{m\}$ and $\{m,m\}$. The first one leads
to one initial context with one location of type $A$ with an instance
of task $m$ in its queue. The second one leads to two contexts, one
with one location of type $A$ with two instances of task $m$ in its
queue, and the other one with two different locations, each with an
instance of task $m$ in its queue.

On the other hand, it makes sense to allow specifying which tasks
should be considered as initial tasks and which should not. A typical
scenario is that the user knows which are the main tasks of the
application and does not want to consider auxiliary or internal tasks
as initial tasks. Another scenario is in the context of integration
testing, where the tester might want to try out together different
groups of tasks to observe how they interfere with each other. Also,
the use of  static analysis can help determine a subset of
tasks of interest to detect some specific property. This is the
case of our deadlock-guided approach of
Section~\ref{sec:initialtasks}.
With all this, the input to our automatic generation of initial
contexts is: (1) a set of \emph{abstract tasks} $\inittasks$ such that
each task is abstracted by the method name that is executing, (2) the
minimum and maximum cardinalities. Thus, an initial context is a set of tuples
(\lst{C.M},$\mincard,\maxcard$), where \lst{C} and \lst{M} are the
class and method name resp., and $\mincard$ resp. $\maxcard$ is the
associated minimum resp. maximum cardinality.
Note that this does not limit the approach in any way since one could
just include in $\inittasks$ all methods in the program and set
$\mincard = 0$ and a sufficiently large $\maxcard$. 


\begin{example}\label{ex:cmincmax}
  Let us consider the set $\inittasks = \sf
  \{(DB.register,1,1),(DB.getData,0,1)\}$. The corresponding
  multisets are $\sf \{register\}$ and $\sf
  \{register,getData\}$. All contexts must contain exactly one
  instance of task \textsf{register} and at most one instance of task
  \textsf{getData}. This leads to three possible contexts: (1) a
  \textsf{DB} location instance with a task \textsf{register} in its queue,
  (2) a \textsf{DB} location instance with tasks \textsf{register} and
  \textsf{getData} in its queue, and, (3) two different \textsf{DB}
  location instances, one of them with an instance of task
  \textsf{register} and the other one with an instance of task
  \textsf{getData}. For instance, the state corresponding to the latter
  context would be: 

\vspace{-.25cm}
\[ \small\begin{array}{llll}
\tt
 S & =\! &\tt [loc(DB1,bot, &\tt [data \mapsto D1, clients
             \mapsto Cl1, checkOn \mapsto B1], \\
   &   &   &\tt [tsk(1,register,[this \mapsto r(DB1), m \mapsto
             W1],body(register))]) \\
   &   &\tt ~loc(DB2,bot, &\tt [data \mapsto D2, clients \mapsto
    Cl2, checkOn \mapsto B2], \\
   &   &   &\tt  [tsk(2,getData,[this
    \mapsto r(DB2), m \mapsto W2],body(getData))])],
\end{array} \]
\noindent where $\sf D1, Cl1,$ and $\sf B1$ (resp. $\sf D2, Cl2,$ and
$\sf B2$) are the fields \textsf{data, clients,} and \textsf{checkOn}
of location $\sf DB1$ (resp. $\sf DB2$), and \textsf{W1}
resp. \textsf{W2} the parameter of the task \textsf{register} resp.
\textsf{getData}, and \textsf{body(m)} is the sequence of instructions
in method \textsf{m}. Note that both fields and task parameters are
fresh variables so that the context is the most general possible.
Note that the first parameter of a task is always the location
\textsf{this} and it is therefore fixed. \hfill $\Box$
\end{example}

In the following, we formally define the contexts that must be
produced from a set of abstract tasks $\inittasks$ with associated
cardinalities, and a procedure (as a Prolog rule) that generates these
contexts as partially instantiated states. 
We use the notation $\{[m_1,...,m_n]_{o_i}\}$ for an initial context
where there exists a location
$\buffer(o_i,\bot,\heap,\{\tkid(tk_1,m_1,\mapping_1,body(m_1))\}\cup...\cup
\{\tkid(tk_n,m_n,\mapping_n,body(m_n))\})$. Note that  we can have $m_i =
m_j$ with $i \neq j$. For instance, the three contexts in
Example~\ref{ex:cmincmax} are written as $\sf
\{[register]_{db_1}\},\{[register,getData]_{db_1}\}$ and $\sf \{[register]_{db_1},[getData]_{db_2}\}$,
respectively.
%
Let us first define the set of initial contexts from a given
$\inittasks$ when all tasks belong to the same class.
\begin{definition}[Superset of initial contexts (same class $C_i$)]
  Let
  $\inittasks=\{(C_i.m_{1},\mincard_1,\maxcard_1),\ldots,(C_i.m_{n},\mincard_n,\maxcard_n)\}$
  be the set of abstract tasks with associated
  cardinalities.  
%
  Let us have $\sum\limits_{i=1}^n\maxcard_i$ different identifiers:
  $o_{1,1},\ldots,o_{1,\maxcard_1}, \\
  \ldots,o_{n,1},\ldots,o_{n,\maxcard_n}$.  We can find at most
  $\sum\limits_{i=1}^n\maxcard_i$ instances of class $C_i$, that is,
  each abstract task $m_{i}$ ($i \in [1,n]$) has at most $\maxcard_i$
  instances and each of them can be inside a different instance of
  class $C_i$.  Let $u^{m_k}_{i,j}$ be an integer variable that
  denotes the number of instances of task $m_k$ inside the location
  $o_{i,j}$ and let us consider the following integer system:
\begin{equation*}
\begin{cases}
\mincard_1 \leq
u^{m_1}_{1,1}+\ldots+u^{m_1}_{1,\maxcard_1}+\ldots+u^{m_1}_{n,1}+\ldots+u^{m_1}_{n,\maxcard_n}
\leq \maxcard_1 \\ 
 \ldots \\
\mincard_n \leq
u^{m_n}_{1,1}+\ldots+u^{m_n}_{1,\maxcard_1}+\ldots+u^{m_n}_{n,1}+\ldots+u^{m_n}_{n,\maxcard_n}
\leq \maxcard_n 
\end{cases} 
\end{equation*}
Each formula requires at least $\mincard_k$ and at most $\maxcard_k$
instances of task $m_k$. Each solution to this system corresponds to
an initial context.
\newline Let
$(d^{m_1}_{1,1},\ldots,d^{m_1}_{n,\maxcard_n},\ldots,d^{m_n}_{1,1},\ldots,d^{m_n}_{n,\maxcard_n})$
be a solution, then the corresponding initial context contains:
\begin{itemize}
\item $\buffer(o_{i,j},\bot,\heap,\queue)$, that is, a location
  $o_{i,j}$ whose lock is free, the fields in $\heap$ are mapped to
  fresh variables, and the queue $\queue$ contains:
  $d^{m_1}_{i,j}$ instances of abstract task $m_1$,\ldots, and
  $d^{m_n}_{i,j}$ instances of $m_n$, if $i \in [1,n]$, $j \in
  [1,\maxcard_i]$ and $\exists d^{m_k}_{i,j} > 0, k \in [1,n]$, where
  each instance of $m_i$ is $tsk(\tkid,m_i,\mapping,body(m_i))$ and
  every argument in $\mapping$ is mapped to a fresh variable.
\end{itemize} 
\end{definition}

\begin{example}\label{ex:superset}
 \noindent Let us consider the example $\inittasks {=} \sf
  \{(DB.register,0,1),(DB.getData,1,1)\}$. The identifiers are
  $o_{1,1}$ and $o_{2,1}$, and the variables of the system are $u^{reg}_{1,1}$, $u^{reg}_{2,1}$,
  $u^{get}_{1,1}$ and $u^{get}_{2,1}$. Finally, we obtain the next system:
\begin{equation*}
\begin{cases}
0 \leq u^{reg}_{1,1} + u^{reg}_{2,1} \leq 1 \\
1 \leq u^{get}_{1,1} + u^{get}_{2,1} \leq 1
\end{cases}
\end{equation*}

\noindent We obtain $6$ solutions: $(0,0,1,0), (0,0,0,1), (1,0,1,0),
(1,0,0,1), (0,1,1,0)$ and $(0,1,0,1)$. Then, the superset of initial
contexts is $$\sf \{\{[getData]_{o_{1,1}}\},\{[getData]_{o_{2,1}}\},
\{[register,getData]_{o_{1,1}}\},\{[register,getData]_{o_{2,1}}\},$$ $$\sf
\{[register]_{o_{2,1}},[getData]_{o_{1,1}}\},\{[register]_{o_{1,1}},[getData]_{o_{2,1}}\}\}  $$ 
 \hfill $\Box$
\end{example}
Let us observe that the two last contexts are equivalent since they
are both composed of two instances of \lst{DB} with tasks
\lst{register} and \lst{getData} respectively. Therefore, we only need
to consider one of these two contexts
for symbolic execution. Considering both would lead to
\emph{redundancy}. The notion of minimal set of initial contexts below
eliminates redundant contexts, hence avoiding useless executions.

\begin{definition}[Equivalence relation $\sim$]
  Two contexts $C_1$ and $C_2$ are equivalent, written $C_1 \sim C_2$,
  if $C_1 = C_2 = \emptyset$ or $C_1 =
  \{\buffer(\objid_1,\bot,\heap_1,\queue_1)\} \cup C_1'$, and
  $\exists~ \objid_2 \in C_2$ such that:
\begin{enumerate} 
\item $C_2 = \{\buffer(\objid_2,\bot,\heap_2,\queue_2)\} \cup C_2'$, 
\item $\queue_1$ and $\queue_2$ contain the same number of instances of
  each task, and
\item $C_1' \sim C_2'$ .
\end{enumerate}
\end{definition}

\begin{example}
  The superset in Example \ref{ex:superset} contains 3 equivalence
  classes induced by the relation $\sim$: (1) the class $\sf
  \{\!\{[getData]_{o_{1,1}}\},\{[getData]_{o_{2,1}}\}\!\},$ where both
  contexts are composed of a location with a task \lst{getData},
  (2) the class $\sf \{\!\{[register,getData]_{o_{1,1}}\}, \\
  \{[register,getData]_{o_{2,1}}\}\!\},$ whose locations have two
  tasks \lst{register} and \lst{getData}.  and, finally, (3) the class
  $\sf
  \{\!\{[register]_{o_{2,1}},[getData]_{o_{1,1}}\},\{[register]_{o_{1,1}},[getData]_{o_{2,1}}\}\!\}
  $, where both contexts have two locations with a task \lst{register}
  and a task \lst{getData}, respectively.  \hfill $\Box$
\end{example}

\begin{definition}[Minimal set of initial contexts
  $\initcontexts{C_i}$ (same class $Cl_i$)]
  Let $\inittasks$ be the set of abstract tasks, then the
  \emph{minimal set of initial contexts} $\initcontexts{Cl_i}$ is
  composed of a representative of each equivalence class induced by
  the relation $\sim$ over the superset of initial contexts for the
  input $\inittasks$.
\end{definition}

\begin{example} 
  As we have seen in the previous example, there are three different
  equivalence classes. So, the minimal set of initial contexts is
  composed of a representative of each class (we have renamed the
  identifiers for the sake of clarity): $$\initcontexts{DB} = \sf
  \{\{[getData]_{{db_1}}\}, \{[register,getData]_{{db_1}}\},
  \{[register]_{{db_1}},[getData]_{{db_2}}\}\} $$ \hfill $\Box$
\end{example}

Let us now define the set of initial contexts $\initcontexts{}$ when
the input set $\inittasks$ contains tasks of different types of
locations.

\begin{definition}[Minimal set of initial contexts  $\initcontexts{}$  (Different classes)]\label{def:different}
  \noindent \\ Let $\inittasks = 
  \{(C_1.m_1,\mincard_1,\maxcard_1),\ldots, (C_n.m_n,\mincard_n,\maxcard_n)\}$
  be the set of abstract tasks with associated cardinalities, and let
  us consider a partition of this set where every equivalence class is
  composed of abstract tasks of the same class. Hence, we have: 
  $\inittasks^{C_1}=\{C_1.m'_1,{\ldots},C_1.m'_{j_1}\},\ldots,\inittasks^{C_n}=\{C_n.m{''}_1,\ldots,C_n.m{''}_{j_n}\}$
  where $C_i \neq C_j, \forall i,j \in [1,n], i\neq j$. \\
Then, 
  let $\initcontexts{C_i}$ be the minimal set of initial contexts for the input
  $\inittasks^{C_i},~ i \in [1,n]$ and $U : \initcontexts{C_1} \times \ldots \times
  \initcontexts{C_n} \rightarrow \initcontexts{}$, defined by
  $U(s_1,\ldots,s_n) = s_1 \cup \ldots \cup s_n$. The set $\initcontexts{}$ is defined by the
  image set of application $U$.
\end{definition} 

\begin{example}\label{ex:initialcontexts}
  Let us consider the set 
  $\inittasks = \sf
  \{(DB.register,1,1),(DB.getData,1,1),\\(Worker.work,1,1)\}$
  from which we get the initial contexts $\initcontexts{Worker}= \sf
  \{\{[work]_{{w_1}}\}\}$ and $\initcontexts{DB} = \sf
\{\{[register,getData]_{{db,1}}\},
\{[register]_{{db_1}},[getData]_{{db_2}}\}\}$. Then, by Def.~\ref{def:different},  
  $$\initcontexts{} \!= \!\sf
  \{ \{[register,getData]_{{db_1}},[work]_{{w_1}}\},\{[register]_{{db_1}},
   [getData]_{{db_2}},[work]_{{w_1}}\}\!\}$$ \hfill $\Box$

\end{example}




\begin{figure}[t]
\begin{center}
\begin{tabular}{ll}
\begin{lstlisting}[style=PrologStyle]
generate_contexts([(M,MinC,MaxC)|Methods],SOut) :-
   add_calls([],[(M,0,MinC,MaxC)|Methods],SOut),
   normal_form(SOut,N),
   (prev_generated(N) -> fail ; assertz(prev_generated(N))).

add_calls(SIn,[(M,Instances,MinC,MaxC)|Ms],SOut) :-
   Instances < MaxC, 
   add_task(SIn,M,SAux),$\label{pp:add-calls12}$
   I2 is Instances + 1, 
   add_calls(SAux,[(M,I2,MinC,MaxC)|Methods],SOut). $\label{pp:add-calls15}$
add_calls(SIn,[(_,I,Min,_),(M,MinC,MaxC)|Methods],SOut) :-
   Min <= I,
   add_calls(SIn,[(M,0,MinC,MaxC)|Methods],SOut). $\label{pp:add-calls22}$
add_calls(SIn,[(_,I,Min,_)],SIn) :-
   Min <= I. $\label{pp:add-calls31}$

add_task([],M,SIn) :-
   fresh_location(LocId),fresh_task(TkId),
   initialize_fields(M,Fields),initialize_mapping(M,L),
   SIn = [loc(LocId,$\bot$,Fields,[tsk(TkId,M,L,body(M))])]. $\label{pp:add-task14}$
add_task([Loc|SIn],M,[Loc2|SIn]) :-
   Loc = loc(Id,Lock,Fields,Q),
   class(Id,Class), class(M,Class), $\label{pp:add-task22}$
   fresh_task(TkId), initialize_mapping(M,L),$\label{pp:add-task23}$
   Loc2 = loc(Id,Lock,Fields,[tsk(TkId,M,L,body(M))|Q]). $\label{pp:add-task24}$
add_task([Loc|SIn],M,[Loc|SOut]) :-
   add_task(SIn,M,SOut). $\label{pp:add-task31}$
\end{lstlisting}
\end{tabular}
\end{center}
\caption{Prolog predicate to generate minimal set of initial contexts}
\label{fig:clp-builtins}
\end{figure}




We now define a Prolog predicate that generates the minimal set of
initial contexts as partially instantiated states. Predicate
\texttt{generate\_contexts/2} in Figure~\ref{fig:clp-builtins}
receives a set of abstract tasks with their associated maximum and
minimum cardinalities, and generates on backtracking all generated
initial contexts by means of \texttt{add\_calls/3}. Predicate
\texttt{normal\_form/2} produces a normal form for the new context
which is the same for all initial contexts in the same equivalence
class. The new context is therefore only generated if it has not been
previously generated (i.e., if the call \texttt{prev\_generated/1}
fails).  The first rule of \texttt{add\_calls/3} checks if the number
of instances \texttt{Instances} of task \texttt{M} is smaller than the
maximum cardinality \texttt{MaxC}, in which case we add a new instance
of \texttt{M}, \texttt{Instances} is incremented, and,
\texttt{add\_calls/3} is recursively invoked. The second rule checks
if the number of instances is greater than or equal to \texttt{Min},
it initializes the number of instances for the next method
(\texttt{M}) and makes the recursive call to
\texttt{add\_calls/3}. Finally, the third rule corresponds to the base
case when we are processing the last method of the list and the number
of instances if greater than or equal to \texttt{Min}.
Predicate \texttt{add\_task/3} adds a new instance of method
\texttt{M} to the current location.  Note here that it can add the new
task to one of the existing locations in \texttt{Locs} or create a new
one to add it.  The first rule checks if \texttt{SIn} is a variable
(the end of the locations list) and then, it creates a new location to
add the task \texttt{M}. To do so, we initialize the location fields,
the method arguments and its tasks queue with a new task with fresh
identifier \texttt{TkId}, and the instructions of method \texttt{M}
(\texttt{body(M)}). The second rule checks if method \texttt{M} can be
added to the first location by checking if the class of location
\texttt{Id} matches with the class of \texttt{M}. If it does, then we
add a new task to its tasks queue \texttt{Q}. The third rule ignores
the first location and tries to add \texttt{M} to \texttt{SIn}.

\begin{example}\label{ex:gen-contexts}
  Let us show predicate \texttt{generate\_contexts/2} in action for
  the set $\inittasks = \sf \{(DB.reg,1,1),(DB.make,1,1), 
  Worker.work,1,1)\}$.  The first rule of \\\texttt{add\_calls/3} is
  applied, as \texttt{0 = Instances < MaxC = 1}. Then,
  \texttt{add\_task/3} is called with variable \texttt{Locs} and
  \texttt{M = DB.register}, at line~\ref{pp:add-calls12}. As
  \texttt{Locs} is a variable, only the first rule of
  \texttt{add\_task/3} can be applied and then, a new location is
  created (line~\ref{pp:add-task14}). Once this predicate has
  finished, \texttt{Instances} is incremented and \texttt{add\_calls}
  is recursively called   (line~\ref{pp:add-calls15}). Now, the second
  rule is applied, as \texttt{0 = Min < Instances = 1},
  and \texttt{add\_calls} is called with \texttt{M = DB.makesConnection}
  whose number of instances is initialized to 0 (line
  \ref{pp:add-calls22}).
  Again, at line\ref{pp:add-calls12}, \texttt{add\_call/3} is
  called with \texttt{M = DB.makesConnection} and \texttt{Locs}
  containing an instance of \lst{DB}. Here we get to a branching point
  which gives rise to the two different initial contexts in
  Example~\ref{ex:initialcontexts}. In the first branch, \texttt{SIn}
  contains a location whose class is equal to that of the method
  \texttt{makesConnection}, so \texttt{LocVar} is the existing
  location and a new instance is added to its queue (lines
  \ref{pp:add-task23} and \ref{pp:add-task24}).
  Finally, \texttt{add\_calls/3} is called with \texttt{M =
    Worker.work} (line \ref{pp:add-calls22}), it creates a new
  instance of class \texttt{Worker} with a task \texttt{work} (line
  \ref{pp:add-task14}), it finishes correctly at line
  \ref{pp:add-calls31}, and returns an initial context containing an
  instance of \lst{DB} with tasks \lst{register} and
  \lst{makesConnection}, and an instance of \lst{Worker} with task
  \lst{work}.
  Now, it fails and the backtracking goes back to the branching
  point. Here, the third rule is applied and then, the first location
  is ignored and task \texttt{makesConnection} is added to a new
  location at line~\ref{pp:add-task14}. It finishes in a similar
  way. In this case, the initial context returned contains two
  instances of \lst{DB} containing a task \lst{register} and
  \lst{makesConnection}, respectively, and an instance of \lst{Worker}
  with task \lst{work}. \hfill $\Box$
\end{example}

\section{On Automatically Inferring Deadlock-Interfering Tasks}\label{sec:initialtasks}

The systematic generation of initial contexts produces a combinatorial
explosion and therefore it should be used with small sets of abstract
tasks (and low cardinalities). However, in the context of deadlock
detection, in order not to miss any deadlock situation, one has to
consider in principle all methods in the program, hence producing
scalability problems. Interestingly, it can happen that many of the
tasks in the generated initial contexts do not affect in any way
deadlock executions.  Our challenge is to only generate initial
contexts from which a deadlock can show up.
For this, the deadlock analysis provides the possibly conflicting task
interactions that can lead to deadlock. We propose to use this
information to help our framework discard initial contexts that cannot
lead to deadlock from the beginning.
Section \ref{sec:deadlock-analysis} summarizes the concepts of the
deadlock analysis used to obtain the deadlock cycles, and Section
\ref{sec:algorithm} presents the algorithm to generate the set of
initial tasks $\inittasks$.

\subsection{Deadlock Analysis and Abstract Deadlock Cycles}
\label{sec:deadlock-analysis}
The deadlock analysis of~\cite{FloresAG13-short} returns a set of
abstract deadlock cycles of the form
$e_1\anndep{\taskpp{p_1}{\abstkid_1}}e_2\anndep{\taskpp{p_2}{\abstkid_2}}...\anndep{\taskpp{p_n}{\abstkid_n}}e_{1}$,
where $p_1,\ldots, p_n$ are program points,
$\abstkid_1,\ldots,\abstkid_n$ are \emph{task abstractions}, and nodes
$e_1,\ldots, e_n$ are either \emph{location abstractions} or task
abstractions. The abstractions
for tasks and locations can be performed at different levels of
accuracy during the analysis: the simple abstraction that we will use
for our formalization abstracts each concrete location $\objid$ by the
program point at which it is created $\objid_{pp}$, and each task by the
method name executing (as in Section~\ref{sec:initialcontexts}). 
They are abstractions since there could be many
locations created at the same program point and many tasks executing
the same method.
 Points-to analysis \cite{Milanova2005,FloresAG13-short}
can be used  to infer such
 abstractions with more precision, for instance, by distinguishing
 the actions performed by  different location abstractions.
 Each arrow $e \anndep{\taskpp{p}{\abstkid}}e'$ should be
 interpreted like ``abstract location or task $e$ is waiting for the
 termination of abstract location or task $e'$ due to the synchronization
 instruction at program point $p$ of abstract task $\abstkid$''.
Three kinds of arrows can be distinguished, namely, \emph{task-task}
(an abstract task is awaiting for the termination of another one),
\emph{task-location} (an abstract task is awaiting for an abstract location to be idle) and
\emph{location-task} (the abstract location is blocked due the
abstract task). \emph{Location-location} arrows cannot happen. 

\begin{example}\label{ex:cycle}
  In our working example there are two abstract locations,
  $\objid_{\ref{pp:newdb}}$, corresponding to location
  \textsf{database} created at line \ref{pp:newdb} and
  $\objid_{\ref{pp:neww}}$, corresponding to the $n$ locations
  \textsf{worker}, created inside the loop at line \ref{pp:neww}; and
  four abstract tasks, $register$, $getD$, $work$ and $ping$.
  The following cycle is inferred by the deadlock analysis:
{\small{$\sf \objid_{\ref{pp:newdb}}\anndep{\ref{pp:register}:register}ping\anndep{\ref{pp:ping}:ping}\objid_{\ref{pp:neww}}\anndep{\ref{pp:work}:work}getD\anndep{\ref{pp:getD}:getD}\objid_{\ref{pp:newdb}}$}}.
%
\noindent The first arrow captures that the location created at
Line \ref{pp:newdb} is blocked waiting for the termination of task
\lst{ping} because of the synchronization at L\ref{pp:register} of task
\lst{register}.  Also, a dependency between a task and a location (for
instance, \lst{ping} and $\sf \objid_{\ref{pp:neww}}$) captures
that the task is trying to execute on that (possibly) blocked
location. Abstract deadlock cycles can be provided by the analyzer to
the user. But, as it can observed, it is complex to figure out from
them why these dependencies arise, and more importantly the interleavings
scheduled to lead to this situation. \hfill $\Box$
\end{example}

\subsection{Generation of initial tasks}\label{sec:algorithm}
The underlying idea is as follows: we select an abstract cycle
detected by the deadlock analysis, and extract a set of potential
abstract tasks which can be involved  in a deadlock.  In a
naive approximation, we could take those abstract tasks that are
inside the cycle and contain a blocking instruction. We also need to
set the maximum cardinality for each task to ensure finiteness (by
default $1$) and require at least one instance for each task (minimum
cardinality).

This approach is valid as long as we only have blocking
synchronization primitives, i.e., when the location state stays
unchanged until the resumption of a suspended execution. 
However, this kind of concurrent/distributed languages usually
include some sort of non-blocking synchronization primitive.  
When a location stops its
execution due to an \Await instruction, another task can
interleave its execution with it, i.e., start
to execute and, thus, modify the location state (i.e., the location
\emph{fields}). 
Then, if a call or a
blocking instruction involved in a deadlock depends on the value of
one of these fields, and we do not consider all the possible values,
a deadlock could be missed. As a consequence, we need to consider
at release points, all possible interleavings with tasks that modify the
fields in order to capture all deadlocks. 

Let us consider now a simple modification of our working
example. Line \ref{pp:connected1} is replaced by \lst{connected = 0}.
Now it is easy to see that if we only consider \lst{register} and
\lst{work} as input, deadlocks are lost: once \lst{register} is
executed and the instruction at line \ref{pp:await} is reached, the
location's queue only contains task \lst{getData} but no \lst{makesConnection} and, therefore, when
task \lst{register} is resumed, field \lst{connected} stays unchanged and
the body of the condition is not executed, so we cannot have a
deadlock situation.

In the following we define the \emph{deadlock-interfering} tasks for a
given abstract deadlock cycle, i.e., an \emph{over-approximation} of
the set of tasks that need to be
considered in initial contexts so that we cannot miss a representative
of the given deadlock cycle. In our extended example, those would be,
\textsf{register} and \textsf{work} but also
\textsf{makesConnection}. 
 
\begin{definition}[initialTasks(C)]\label{def:initialtasks}\small
Let C an abstract deadlock cycle. Then, 
$$initialTasks(C) := \bigcup\limits_{i_{call} \in t \in C}
\!\!\!initialTasks(t,i_{call},C) ~\cup~ \bigcup\limits_{i_{sync} \in t \in C} \!\!\!initialTasks(t,i_{sync},C)$$

where: 
\begin{itemize}
\item $initialTasks(t,i,C) = \emptyset   ~~~~~if~ \objid
\anndep{t} t_2 \not \in C ~and~ i \neq i_{mod}~and~{\not \exists}~i_{await}~{\in}
[t_0,i]$

\item $initialTasks(t,i,C) = \{t\}   ~if~ (\objid \anndep{t} t_2 \in C
~or~ i =  i_{mod}~) ~and~ {\not \exists}~i_{await}~{\in} [t_0,i]$

\item $initialTasks(t,i,C) = \{t\} ~\cup \bigcup\limits_{f \in fields(i)}
\left(\bigcup \limits_{i_{mod} \in t_{mod} \in mods(f)}
\!\!\!\!\!\!\!\!\!\!initialTasks(t_{mod},i_{mod},C) \right) \\ \text{~~~~~~~~~~~~~~~~~~~~~~~~~~~~~~~~}~if~ \exists~i_{await}~\in
  [t_0,i]$
\end{itemize}
\end{definition}
\normalsize

The definition relies on function \textsf{fields(I)} which, given an
instruction \textsf{I}, returns the set of class fields that have been
read or written until the execution of instruction \textsf{I}. Let
\textsf{mods(f)} be the set of instructions that modify field
\textsf{f}.  We can observe that \emph{initialTasks(C)} is the union
of initial tasks for each relevant instruction inside the cycle C,
i.e., asynchronous calls and synchronization primitives. We can
also observe in the auxiliary function \emph{initialTasks(t,i,C)} that:
(1) if the instruction $i$ is not producing a \emph{location-task
  edge} and it is not an instruction modifying a field, then $t$ does
not  need to be added as initial task, (2) if $i$ produces a
\emph{location-task edge} or is modifying a field, and we  do not
have any \Await instruction between the beginning of the task and
$i$, then $i$ is going to be executed under the most general context,
so we do not need to add more initial tasks but $t$, and (3) on the
other hand, if there exists an \Await instruction between the
beginning of task $t$, namely $t_0$, and instruction $i$, each field $f$ inside the set
\textsf{fields(i)} could be changed before the resumption of the \Await by
any task modifying $f$. Thus, tasks containing any of the possible
$f$-modifying instructions must be considered and, recursively, their
initial tasks.

It is important to highlight that this definition could be infinite
depending on the program we are working with. For instance, if we
apply the definition to the abstract cycle $C$ in Example
\ref{ex:cycle},
$initialTasks(\text{\textsf{db.register}},\ref{pp:connected},C)$ will
be evaluated. It fits well with the conditions on third clause, as
there exists an \Await~ instruction,
\textsf{fields(\ref{pp:connected})} = \{\textsf{connected}\} and then
again \ref{pp:connected} is a modifier instruction of field
\textsf{connected}, so
$initialTasks(\text{\textsf{db.register}},\ref{pp:connected},C)$ will
be evaluated again recursively.

\begin{figure}[t]
\begin{center}
\begin{tabular}{ll}
\begin{lstlisting}[style=PrologStyle]
calculate_interfering_tasks(Cycle,Tasks) :-
   init(Cycle,[],Events,[],Ans),
   process_events(Events,Ans,NoCardinality),
   findall((Task,1,1),member((Task,_),NoCardinality),Repeated),
   list_to_set(Repeated,Tasks).
 
init([],Evs,Evs,Ans,Ans).
init([edge(loc,get(Task,LAsync,LGet),task)|C],Evs,Evs2,Ans,Ans2) :-
   !, init(C,[(Task,LAsync)$\!$,$\!$(Task,LGet)|Evs]$\!$,Evs2,[(Task,LGet)|Ans],Ans2)$\!$.
init([edge(task,sync(Task,LAsync,LSync),task)|C],Evs,Evs2,Ans,Ans2) :-
   !, init(C,[(Task,LAsync),(Task,LSync)|Evs],Evs2,Ans,Ans2).
init([_|Cycle],Evs,Evs2,Ans,Ans2) :- init(Cycle,Evs,Evs2,Ans,Ans2).

process_events([],Ans,Ans).
process_events([(Task,Inst)|Evs],Ans,Ans2) :-
   thereis_await(Task,Inst),
   accessed_fields(Task,Inst,Fields), !,
   findall((T,L),(member(F,Fields),
                    inst(F,write,T,L),
                    \+ member((T,L),Ans)),Modifiers),
   append(Modifiers,Evs,Evs2), append(Modifiers,Ans,Ans1),
   process_events(Evs2,Ans1,Ans2).
process_events([_|Evs],Ans,Ans2) :- process_events(Evs,Ans,Ans2).
\end{lstlisting}
\end{tabular}
\end{center}
\caption{Prolog predicate to infer interfering tasks for a given
  deadlock cycle}
\label{alg:tasks2}
\end{figure}

Figure \ref{alg:tasks2} presents predicate
\texttt{calculate\_interfering\_tasks/2} that finitely infers the
interfering-tasks for a given deadlock cycle as defined by
Def~\ref{def:initialtasks}. 
First, both the list of events and of answers are initialized
(\texttt{init/5}) according to the type of edge. For each edge in the
cycle, we take the call and the corresponding synchronization
instruction, and we add them to the pending events. Moreover, \Get\
instructions produce \emph{location-task} edges, so they are also
included in the answers list, as they have to be inside the initial
context. The other tasks included in the initial context are the ones
which could affect the conditions of those instructions.
In predicate \texttt{process\_events/3}, we take a pending event
\texttt{(Task,Inst)} and we check if there is an \Await
instruction between the start of \texttt{Task} and \texttt{Inst}, using
predicate 
(\texttt{thereis\_await/2}), where the previously accessed field values
(\texttt{accessed\_fields/3}) could be changed (third clause in
Def.~\ref{def:initialtasks}. In case it does, we need to include in
the answer set all tasks which contain instructions modifying such
field (\texttt{inst/4}). Besides, this change could be inside an
if-else body and we also need to consider the fields inside such
condition. Therefore we add the modifier instructions to the pending
events list. This predicate finishes when this list is empty and \texttt{Ans}
is the list of pairs with all interfering instructions and their
container tasks. Finally, we only take the tasks, i.e., the first
component of each pair, we set their minimum and maximum cardinalities
and remove duplicates (\texttt{list\_to\_set/2}). Finiteness is
guaranteed because each instruction is added to the pending events and
answers lists at most once, and the number of instructions is finite.

\begin{example}
  Let us show how predicate \texttt{calculate\_interfering\_tasks/2} works for our modified example. For
  the sake of clarity, instructions are identified by their line
  numbers. After the \texttt{init/5} predicate, the value of variables
  $\tt Events$ and $\tt Ans$ is $\sf
  [(Worker.work,\ref{pp:work}),(Worker.work,\ref{pp:fgetdata}),
  (DB.register,\ref{pp:register}),(DB.register,\ref{pp:fping})]$ and
  $\sf [(DB.register,\ref{pp:register}),(Worker.work,\ref{pp:work})]$,
  respectively. Hence, predicate \texttt{process\_e\-vents/3} takes $\sf
  (Worker.work,\ref{pp:work})$ first. Since there is not an
  \Await instruction between the beginning of \lst{work} and line \ref{pp:work},
  {\tt Ans} stays unchanged.  The
  same happens with $\sf (Worker.work,\ref{pp:fgetdata})$. Now, the
  pending events list is $\sf
  [({DB.register},\ref{pp:register}),$ $\sf
  ({DB.register},\ref{pp:fping},)]$ and $\sf
  ({DB.register},\ref{pp:register})$ is processed. Now, there is
  an \Await between lines \ref{pp:reginit} and \ref{pp:register} and, then,
  \texttt{fields({\sf DB.register},\ref{pp:register},Fields)} is invoked and
  $\tt Fields= \sf [connected] $.  We find three instructions
  modifying the field \lst{connected}: \ref{pp:makestrue} $\in$
  \lst{DB.makesConnection}, \ref{pp:connected1} $\in$
  \lst{DB.register} and \ref{pp:connected} $\in$
  \lst{DB.register}. None of them is a member of the answer set and hence they
  are added to both lists. Now, $\tt Evs$ is \textsf{
  [({DB.register},\ref{pp:connected}),
  ({DB.makesConnection},\ref{pp:makestrue}),
  ({DB.register},\ref{pp:fping}),({DB.register},\ref{pp:connected1})]}
  but again there is no \Await between the beginning of
tasks {\sf DB.register} and {\sf DB.makesConnection} and lines \ref{pp:connected} and \ref{pp:makestrue},
respectively and, thus, $\tt Ans$
  stays unchanged.  Finally, both $\sf ({DB.register},\ref{pp:connected1})$
  and $\sf ({DB.register},\ref{pp:fping})$ are taken and both 
  \texttt{fields({\sf DB.register},\ref{pp:fping},Fields)} and
  \texttt{fields({\sf DB.register},\ref{pp:connected1},Fi\-elds)} hold
  where $\tt
  Fields {=} \sf [connected]$, but the
  modifier instructions have been previously added to $\tt Ans$, hence
  $\tt Ans$
  remains unchanged, and the pending events list becomes empty.
  Finally, the algorithm
  projects over the first component of each pair in the list, sets the
  minimum and maximum cardinalities to $1$ and removes
  duplicates, returning the set $\inittasks =$ \textsf{
  \{{(DB.register,1,1)}, {(Worker.work,1,1)},
  {{(DB.ma\-kes\-Connection,1,1)}}\}}. Thus, the generation of initial contexts
  for this set (see Example~\ref{ex:gen-contexts}) produces
\[
\begin{array}{lll}
  \initcontexts{} &=\{&
  {\sf\{[register,makesConnection]_{db_1}[work]_{w_1}\},}\\
  &&{\sf\{[register]_{db_1},[makesConnection]_{db_2},[work]_{w_1}\}\}}
\end{array}
\]
  \hfill $\Box$
\end{example}



\section{Conclusions and Related Work}\label{sec:concl-future-work}


We have proposed a framework for the automatic generation of initial
contexts for deadlock-guided symbolic execution. Such initial contexts
are composed of the interfering tasks which, according to a static
deadlock analyzer, might lead to deadlock. Given the initial contexts,
we can drive symbolic execution towards paths that are more likely to
manifest a deadlock, discarding safe contexts.
There is a large body of work on deadlock detection including both
dynamic and static approaches.  Much of the existing work, both for
asynchronous programs \cite{FloresAG13-short,GGLLW13} and
thread-based programs
\cite{DBLP:conf/pdd/MasticolaR91,DBLP:journals/tocs/SavageBNSA97}, is
based on static analysis techniques. Although we have used the static
analysis of \cite{FloresAG13-short}, the information provided by other
deadlock analyzers could be used in an analogous way.
Deadlock detection has been also studied in the context of dynamic
testing and model checking
\cite{DBLP:conf/icst/ChristakisGS13,DBLP:conf/pldi/JoshiPSN09,lockout},
where sometimes has been combined with static information
\cite{DBLP:conf/hvc/AgarwalWS05,DBLP:conf/sigsoft/JoshiNSG10}. 
The initial contexts generated by our framework are of interest also
in these approaches. Deadlock detection is even more challenging in
the context of thread-based concurrency model. As future work, we plan to
investigate how our framework could be adapted to this model.





\bibliographystyle{plain}


\end{document}